\documentclass[preprint,showkeys,secnumarabic,amsfonts,amsmath,amssymb]{revtex4}
\usepackage[dvips]{color}
\usepackage{array}
\usepackage{epsfig}
\usepackage{amsmath}
\usepackage{graphicx}

\begin{document}

\title{Logamediate inflation in DGP cosmology driven by a non-canonical scalar field}

\author{A. Ravanpak}
\email{a.ravanpak@vru.ac.ir}
\affiliation{Department of Physics, Vali-e-Asr University of Rafsanjan, Rafsanjan, Iran}

\author{G. F. Fadakar}
\email{g.farpour@vru.ac.ir}
\affiliation{Department of Physics, Vali-e-Asr University of Rafsanjan, Rafsanjan, Iran}

\date{\today}

\begin{abstract}

The main properties of the logamediate inflation driven by a non-canonical scalar field in the framework of DGP braneworld gravity are investigated. Considering high energy conditions we calculate the slow-roll parameters, analytically. Then, we deal with the perturbation theory and calculate the most important respective parameters such as the scalar spectral index and the tensor-to-scalar ratio. We find that the spectrum of scalar fluctuations is always red-tilted. Also, we understand that the running in the scalar spectral index is nearly zero. Finally, we compare this inflationary scenario with the latest observational results from Planck 2018.

\end{abstract}

\keywords{logamediate inflation, tachyon, DGP, perturbation, Planck}

\maketitle

\section{Introduction}

Inflation is a very short period at extremely early stages of the history of the Universe in which the Universe experiences a very rapidly accelerated expansion \cite{Guth},\cite{Albrecht}. It is one of the most compelling solutions to many long-standing problems of the standard hot big bang scenario such as the flatness problem and the horizon problem. Besides, it has provided a causal interpretation of the origin of the anisotropies in the cosmic microwave background radiation, and also the distribution of large scale structure of the Universe \cite{Liddle}-\cite{Hinshaw}.

There are various inflationary models with different scale factor parameter $a(t)$, such as the standard de-Sitter inflation with $a(t)=\exp(Ht)$ in which $H=\dot a/a$ (dot means derivative with respect to the cosmic time, $t$) is the Hubble parameter, the power law inflation with $a(t)=t^p$ in which $p>1$, the intermediate inflation with $a(t)=\exp(At^f)$ in which $A>0$ and $0<f<1$ and so on. There is another class of inflationary models called logamediate inflation that naturally arises from some scalar-tensor theories in which the scale factor behaves as \cite{Barrow}
\begin{equation}\label{scalefactor}
    a(t)=\exp(A[\ln t]^\lambda)
\end{equation}
Here, $A>0$ and $\lambda>1$ are two constants. Logamediate inflation can also be extracted from a new class of cosmological solutions with an indefinite expansion that in turn results when weak general conditions apply on the cosmological models \cite{Barrow2}. Although logamediate inflation belongs to the class of non-oscillating inflationary models in which the inflation does not naturally terminate, one can use various approaches such as the curvaton scenario to bring inflation to an end \cite{Farajollahi},\cite{Farajollahi2}.

In general, inflation drives by the potential of a standard canonical scalar field called inflaton. But, there is a non-canonical scalar field named tachyon motivated from string theory that is responsible to drive an inflationary phase \cite{Sen}-\cite{Sami}. The tachyonic inflation is a special kind of $k$-inflationary models \cite{Picon} in which the tachyon field $\phi$ that has a positive potential $V(\phi)$ starts its evolution from an unstable maximum at $\phi=0$ and finally, when $|\phi|\rightarrow\infty$, $V(\phi)\rightarrow0$ while $V'=dV/d\phi$ is always negative.

The idea of a tachyon logamediate inflation has been studied in \cite{Ravanpak} in which the authors have found good agreements between their model and the observational data from seven-year Wilkinson Microwave Anisotropy Probe. But, higher dimensional theories of gravity in which the standard model of particle physics is restricted to a four-dimensional (4D) brane and only gravity can leak into the higher dimensional space called bulk may explain the inflationary phase more accurately, because the extra dimensions induce additional terms into the Friedmann equation on the brane. Brane inflationary scenarios have been widely studied in the literature \cite{Setare}-\cite{Maartens}. Especially, in \cite{Kamali} a logamediate inflationary model driven by a tachyon field has been investigated in the context of brane cosmology. The authors have indicated the consistency between their model and Planck 2013 results.

A modified braneworld scenario was put forward by Dvali, Gabadadze and Porrati (DGP) in which a 4D scalar curvature term added to the matter Lagrangian on the brane and the bulk considered to be Minkowskian \cite{Dvali}. Depending on how the brane can be embedded into the five-dimensional (5D) bulk, the DGP model results in two different solutions. The self-accelerating solution denoted by $\epsilon=+1$ that produces the late-time acceleration of the Universe naturally but suffers from the ghost instability and the normal solution denoted by $\epsilon=-1$ that needs a dark energy component to explain the late-time acceleration. In spite of the merit of the DGP scenario in explaining the late-time acceleration of the Universe, many efforts have been devoted to study the DGP inflationary models \cite{Herrera}-\cite{Cai}. Especially, in \cite{Ravanpak2} the authors have studied an intermediate inflationary scenario in the DGP cosmology driven by a tachyon scalar field and have constrained their model parameters using the latest observational results borrowed from Planck 2015.

In this article we investigate a tachyon logamediate inflation in the DGP braneworld cosmology. The structure of the article is as follows: In the next section we begin with the action of the DGP model. Considering the slow-roll inflationary conditions and in the high energy regime we obtain the effective Friedmann equation, the evolutionary equation of the tachyon scalar field and the slow-roll parameters. In Sec.\ref{sec3}, we deal with perturbation theory and calculate respective parameters. Numerical discussions are brought up in Sec.\ref{sec4}, as well as some analytical deductions. To find if our model is reliable or not, we use recent observational constraints from Planck satellite. Finally, the last section will demonstrate a summary of the work and the main results.

\section{THE MODEL}\label{sec2}

We start with the action of the DGP braneworld model as follows:
\begin{equation}\label{action}
S=\frac{1}{2\kappa_5^2}\int d^5x\sqrt{-g^{(5)}}R^{(5)}+\int d^4x\sqrt{-g} \cal L
\end{equation}
in which the first and the second term demonstrate a 5D Minkowskian Einstein-Hilbert action and the contribution of the induced gravity localized on the brane, respectively. Also, $g^{(5)}$, $R^{(5)}$ and $g$ are the 5D metric, the 5D Ricci scalar and the induced 4D metric on the brane, individually. $\cal L$ indicates the effective 4D Lagrangian on the
brane which can be expressed as
\begin{equation}\label{Lagrangian}
{\cal L} =\frac{\mu^2}{16\pi}R+L_m
\end{equation}
in which $R$ and $L_m$ are the 4D Ricci scalar and the matter Lagrangian on the brane, respectively. Also, $\mu$ is a mass parameter which controls the power of the induced gravity term and may correspond to the 4D Planck mass $m_p$. Obviously, in our model the bulk is considered to be empty and all the matter content are confined into the brane.

Considering a spatially flat FRW metric on the brane and introducing $\rho_0=\frac{48\pi}{\kappa_5^4\mu^2}$ we obtain the Friedmann equation on the brane as follows:
\begin{equation}\label{Friedmann}
H^2=\frac{8\pi}{3\mu^2}(\sqrt{\rho+\frac{\rho_0}{2}}+\epsilon\sqrt{\frac{\rho_0}{2}})^2
\end{equation}
in which $\rho$ is the matter density on the brane that obeys the standard conservation equation as
\begin{equation}\label{conservation}
\dot \rho+3H(\rho+p)=0
\end{equation}
and $p$ indicates the pressure.

Since we are interested in an inflationary period that relates to the early Universe we can apply the high energy approximation $\rho\gg\rho_0$ (\cite{Lopez},\cite{Maeda}) into Eq.(\ref{Friedmann}) to obtain the effective Friedmann equation on the brane in the inflationary era as
\begin{equation}\label{effectiveFriedmann}
H^2=\frac{8\pi}{3\mu^2}(\sqrt{\rho}-\sqrt{\frac{\rho_0}{2}})^2
\end{equation}
in which we have set $\epsilon=-1$ that means we have considered a normal branch of the DGP model. Also, if we assume that in this period a tachyon scalar field that is responsible for the inflation is the dominant matter component on the brane we replace the energy density and the pressure of a tachyon scalar field
\begin{equation}\label{energydensity}
\rho=\frac{V(\phi)}{\sqrt{1-\dot \phi^2}}, \quad p=-V(\phi)\sqrt{1-\dot \phi^2}
\end{equation}
into Eq.(\ref{conservation}) to obtain the tachyon evolutionary equation as follows:
\begin{equation}\label{EoM}
\frac{\ddot \phi}{1-\dot \phi^2}+3H\dot\phi+\frac{V'}{V}=0
\end{equation}
Using the Eqs.(\ref{conservation}), (\ref{effectiveFriedmann}), (\ref{energydensity}) and (\ref{EoM}) we obtain
\begin{equation}\label{phidot2}
\dot\phi^2=-\frac{2\dot H}{3H^2(1+\alpha/H)}
\end{equation}
in which $\alpha=\sqrt{\frac{4\pi\rho_0}{3\mu^2}}$. For logamediate inflationary scenario with the scale factor as Eq.(\ref{scalefactor}), we find that
\begin{equation}\label{phi}
\phi(t)=\int\sqrt{\frac{2(\ln t-\lambda+1)}{3[A\lambda(\ln t)^\lambda+\alpha t\ln t]}}dt
\end{equation}
Also, using the Eqs.(\ref{effectiveFriedmann}), (\ref{energydensity}) and (\ref{phidot2}) one can obtain an effective potential for the model as
\begin{equation}\label{effectivepotential}
V=\frac{3\mu^2H^2}{8\pi}\left(1+\frac{\alpha}{H}\right)^2\sqrt{1+\frac{2\dot H}{3H^2(1+\alpha/H)}}
\end{equation}

Besides, to drive a long enough inflation the tachyon scalar field must slowly rolls down its potential. In this mechanism called slow-roll inflation the energy density of the
inflaton field and its potential must satisfy $\rho\sim V(\phi)$. In the case of a tachyonic inflation the slow-roll conditions are $\dot\phi^2\ll1$ and $\ddot\phi\ll3H\dot\phi$. So, Eqs.(\ref{effectiveFriedmann}), (\ref{EoM}) and (\ref{effectivepotential}) reduce to
\begin{equation}\label{SReffectiveFriedmann}
H^2\approx\frac{8\pi}{3\mu^2}(\sqrt{V}-\sqrt{\frac{\rho_0}{2}})^2,
\end{equation}
\begin{equation}\label{SREoM}
\frac{V'}{V}\approx-3H\dot\phi
\end{equation}
and
\begin{equation}\label{SReffectivepotential}
V\approx\frac{3\mu^2H^2}{8\pi}\left(1+\frac{\alpha}{H}\right)^2
\end{equation}
respectively.

Slow-roll parameters are some dimensionless parameters that are defined in slow-roll inflationary models. In terms of our model parameters they can be written as
\begin{equation}\label{epsilon}
\varepsilon=-\frac{\dot H}{H^2}=\frac{(\ln t)^{-\lambda}}{A\lambda}(\ln t-\lambda+1)
\end{equation}
and
\begin{equation}\label{eta}
\eta=-\frac{\ddot H}{H\dot H}=\frac{(\ln t)^{-\lambda}}{A\lambda}\left[2\ln t-\lambda+2+\frac{\ln t}{\lambda-1-\ln t}\right]
\end{equation}
The general behavior of $\varepsilon$ in Eq.(\ref{epsilon}) is as follows: it starts to increase at $t=1$ and keeps increasing until it reaches to a maximum $\varepsilon_{max}$ at a given time $t_{max}$. Then, it returns and approaches zero as $t\rightarrow\infty$. Besides, we know that the inflationary phase takes place for $\ddot a>0$ which in turn is proportional to $\varepsilon<1$, so that at the beginning or at the end of inflation we have $\varepsilon=1$. Therefore, we study only the situations in which $\varepsilon_{max}\geq1$ and consider $\varepsilon=1$ at a given time $t_b\geq t_{max}$ as the beginning of inflation and leave $t<t_b$ for the pre-inflationary Universe. With attention to $\varepsilon_{max}\geq1$, it is easy to calculate a constraint as follows:
\begin{equation}\label{constraint}
A\leq\lambda^{-\lambda-1}
\end{equation}
Also, to find $t_b$ we consider another approximation as $\ln t\gg\lambda-1$ that is valid at late times \cite{Barrow3}. Using this approximation Eq.(\ref{epsilon}) reduces to $\varepsilon=(\ln t)^{1-\lambda}/(A\lambda)$ which yields
\begin{equation}\label{tb}
t_b=\exp\left[(A\lambda)^{\frac{1}{1-\lambda}}\right]
\end{equation}
Also, one can check  that using this approximation, Eq.(\ref{eta}) reduces to $\eta=2\varepsilon$.

The number of $e$-folds between two cosmological times $t_1$ and $t_2>t_1$ in our model is obtained as below
\begin{equation}\label{N}
N=\int_{t_1}^{t_2} H dt=A[(\ln t_2)^{\lambda}-(\ln t_1)^{\lambda}]
\end{equation}
Considering $t_1=t_b$, one can find that
\begin{equation}\label{lnt}
\ln t=\left[\frac{N}{A}+(A\lambda)^{\frac{\lambda}{1-\lambda}}\right]^{\frac{1}{\lambda}}
\end{equation}
that is a useful equation in the following.

\section{PERTURBATION}\label{sec3}

Recent observations have revealed small deviations from an entirely homogeneous Universe and we believe that gravity has caused very tiny inhomogeneities of the early Universe to grow with time because of its attractive feature. So, we can use the linear perturbation theory to have a more exactly understanding of our Universe. To this aim we use the general linearly perturbed flat Friedmann-Robertson-Walker metric that includes both scalar and tensor perturbations. The power spectrum of the curvature perturbation $\cal{P_R}$ is one of the most important parameters in the perturbation theory that for the tachyon field is defined as ${\cal{P_R}}=(\frac{H^2}{2\pi\dot\phi})^2\frac{1}{Z_s}$ in which $Z_s=V(1-\dot\phi^2)^{-3/2}$ \cite{Hwang}. Note that this relation is just the result obtained in a standard 4D gravity for a tachyon scalar field, because it is a consequence of the local conservation of energy-momentum in four dimensions and is independent of the form of the gravitational field equations \cite{Wands}. Moreover, in the case of slow-roll inflation it reduces to ${\cal{P_R}}=(\frac{H^2}{2\pi\dot\phi})^2\frac{1}{V}$ that with attention to Eqs.(\ref{phidot2}) and (\ref{SReffectivepotential}) can be written in our model as
\begin{equation}\label{PR}
{\cal{P_R}}\approx\frac{H^2}{\pi\mu^2\varepsilon(1+\alpha/H)}
\end{equation}
Considering logamediate inflation the above relation can be rewritten as
\begin{equation}\label{PRt}
{\cal{P_R}}\approx\frac{A^4\lambda^4(\ln t)^{4\lambda-4}}{\pi\mu^2t^2[A\lambda(\ln t)^{\lambda-1}+\alpha t]}
\end{equation}

Another important parameter in the perturbation theory is the scalar spectral index $n_s$ which corresponds to the power spectrum of the scalar perturbations via the relation $n_s-1=d\ln{\cal{P_R}}/d\ln k$ in which the interval in wave number $k$ is related to the number of $e$-folds through $d\ln k=dN$ \cite{Campo3}. One can check that in our model the scalar spectral index can be written as follows:
\begin{equation}\label{ns}
n_s\approx1-\varepsilon\left(2+\frac{\frac{\alpha}{H}}{1+\frac{\alpha}{H}}\right)
\end{equation}
The above relation can be rewritten in terms of $N$, using Eq.(\ref{lnt}) (which in turn comes from the late time approximation) as:
\begin{equation}\label{nsN}
n_s\approx1-\frac{{\cal N}^{\frac{1-\lambda}{\lambda}}}{A\lambda}
\left[2+\frac{\alpha\exp\left({{\cal N}^{\frac{1}{\lambda}}}\right)}
{A\lambda{\cal N}^{\frac{\lambda-1}{\lambda}}
+\alpha\exp\left({{\cal N}^{\frac{1}{\lambda}}}\right)}\right]
\end{equation}
in which we have replaced $\frac{N}{A}+(A\lambda)^{\frac{\lambda}{1-\lambda}}$ with $\cal N$.

The third interesting parameter that is strongly related to $n_s$, is the running in the scalar spectral index indicated by $n_{run}$. It can be calculated via the relation $n_{run}=dn_s/d\ln k$. We would like to note that if $n_s$ does not depend on scale then $n_{run}=0$. In our model we obtain
\begin{equation}\label{nrun}
n_{run}\approx-\varepsilon^2\left(\frac{\frac{\alpha}{H}}{[1+\frac{\alpha}{H}]^2}\right)
\end{equation}
Again, in terms of $N$ we can rewrite the prior equation as below
\begin{equation}\label{nrunN}
n_{run}\approx-\frac{{\cal N}^{\frac{2(1-\lambda)}{\lambda}}}{A^2\lambda^2}
\left[\frac{\alpha A\lambda{\cal N}^{\frac{\lambda-1}{\lambda}}\exp\left({{\cal N}^{\frac{1}{\lambda}}}\right)}
{\left[A\lambda{\cal N}^{\frac{\lambda-1}{\lambda}}
+\alpha\exp\left({{\cal N}^{\frac{1}{\lambda}}}\right)\right]^2}\right]
\end{equation}

Besides, the generation of tensor perturbations during inflation would produce gravitational waves and these perturbations are more involved in our model, since in the braneworld cosmology the gravitons can leak into the bulk. The amplitude of tensor perturbations in an induced gravity model is given by \cite{Lopez2},\cite{Lidsey}
\begin{equation}\label{Pg}
{\cal{P}}_g=\frac{64\pi}{m_p^2}\left(\frac{H}{2\pi}\right)^2G_\gamma^2(x)
\end{equation}
in which $G_\gamma^{-2}(x)=\gamma+(1-\gamma)F^{-2}(x)$ and $\gamma=\mu^2/m_p^2$. It is clear that in one hand the above equation reduces to the form of the power spectrum of tensor perturbations in a standard 4D cosmology for $\gamma=1$ and on the other hand to the one in a braneworld scenario without an induced gravity correction such as the Randall-Sundrum model for $\gamma\rightarrow0$. Also, in Eq.(\ref{Pg}), $F(x)$ is defined as
\begin{equation}\label{Fx}
F(x)=\left[\sqrt{1+x^2}-x^2\sinh^{-1}(\frac{1}{x})\right]^{-\frac{1}{2}}
\end{equation}
Here, $x=H/\bar{\mu}$ in which $\bar{\mu}$ denotes the energy scale associated with the bulk curvature. As we mentioned earlier the bulk is Minkowskian in the DGP braneworld cosmology, so that $\bar{\mu}\rightarrow0$. In other words the energy scale at which inflation begins will satisfies the condition $H\gg\bar{\mu}$. As it has been discussed in \cite{Lopez2} under this condition Eq.(\ref{Pg}) reduces to
\begin{equation}\label{Pgreduced}
{\cal{P}}_g\approx\frac{64\pi}{m_p^2}\left(\frac{H}{2\pi}\right)^2\frac{1}{\gamma}
\end{equation}
that for a logamediate inflationary scenario can be written as
\begin{equation}\label{Pgt}
{\cal{P}}_g\approx\frac{16A^2\lambda^2(\ln t)^{2\lambda-2}}{\pi\mu^2t^2}
\end{equation}

Also, in a power-law parametrization for tensors another parameter appears called tensor spectral index $n_T$ which is defined as $n_T=d\ln{\cal{P}}_g/d\ln k$ and in our model is obtained as follows:
\begin{equation}\label{nT}
n_T\approx-2\varepsilon
\end{equation}

Another important quantity in perturbation theory is the so called tensor-to-scalar ratio $r$ which is the ratio between the power spectrum of the tensor perturbations and the power spectrum of the scalar perturbations. In our model it can be obtained as follows:
\begin{equation}\label{r}
r=\frac{{\cal{P}}_g}{{\cal{P_R}}}\approx16\varepsilon\left(1+\frac{\alpha}{H}\right)
\end{equation}
in which we have used Eqs.(\ref{PR}) and (\ref{Pgreduced}). This relation can be rewritten in terms of the number of $e$-folds using Eq.(\ref{lnt}), as follows:
\begin{equation}\label{rN}
r\approx\frac{16\left[A\lambda{\cal N}^{\frac{\lambda-1}{\lambda}}+\alpha \exp\left({\cal N}^{\frac{1}{\lambda}}\right)
\right]}{A^2\lambda^2{\cal N}^{\frac{2\lambda-2}{\lambda}}}
\end{equation}
Using Eqs.(\ref{nT}) and (\ref{r}), one can calculate the consistency relation between the two parameters tensor to scalar ratio and tensor spectral index as
\begin{equation}\label{rnT}
r=-8n_T(1+\frac{\alpha}{H})
\end{equation}
which differ form the one in the case of slow-roll inflation driven by a single canonical scalar field by the factor $1+\alpha/H$.

In the following we try to check the consistency of the model with observations, numerically.

\section{numerical discussion}\label{sec4}

\begin{figure}[h]
\centering
\includegraphics[width=8cm]{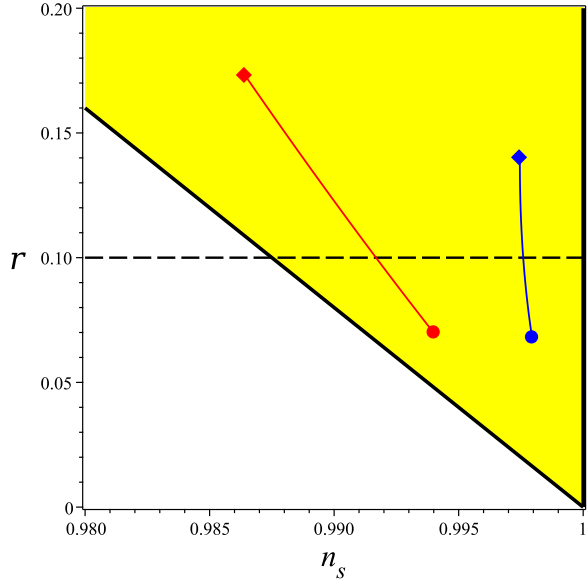}
\caption{All the possible trajectories of the model in $(r-n_s)$ plane lie in the allowed yellow region between the curves $n_s=1$ and $r=8(1-n_s)$. The red (left) curve and the blue (right) curve relates to ($\lambda$=10, $A=10^{-11}$) and ($\lambda$=20, $A=4.7\times10^{-28}$), respectively. In both the cases we have assumed $\alpha=2\times10^{-12}$. Also, in the red (left) curve the diamond and the circle indicates the number of $e$-fold parameter $N=100$ and $N=300$, while in the blue (right) curve the diamond and the circle indicates $N=60$ and $N=150$, respectively. The horizontal black dashed line represents the recent constraint on the tensor-to-scalar ratio parameter, $r<0.1$, by Planck collaboration \cite{Akrami}.}\label{fig1}
\end{figure}

In the study of perturbations, $(r-n_s)$ diagram may be so useful. For the model under consideration one can obtain a relation between $r$ and $n_s$, using Eqs.(\ref{ns}) and (\ref{r}) as follows:
\begin{equation}\label{r-ns}
r=\frac{16(1+\frac{\alpha}{H})(1-n_s)}{2+\left(\frac{\frac{\alpha}{H}}{1+\frac{\alpha}{H}}\right)}
\end{equation}
Obviously, this relation depends on the values of the model parameters $A$, $\lambda$ and $\alpha$ in addition to $N$. But we can find two boundary trajectories for two cases $\alpha\ll H$ and $\alpha\gg H$ that are independent of the other parameters' values because $H$ slowly decreases during logamediate inflation while $\alpha$ is a constant. In one hand for $\alpha\ll H$, we reach to $r\approx8(1-n_s)$ and on the other hand for $\alpha\gg H$, we find the Harrison-Zel'dovich model $n_s\approx1$. Thus using Eqs.(\ref{nsN}) and (\ref{rN}), all the possible trajectories in $(r-n_s)$ plane for a set of $(A,\lambda,\alpha)$ locate somewhere between them. The allowed region in $(r-n_s)$ plane has been shown in FIG.\ref{fig1}.

\begin{figure}[h]
\centering
\includegraphics[width=8cm]{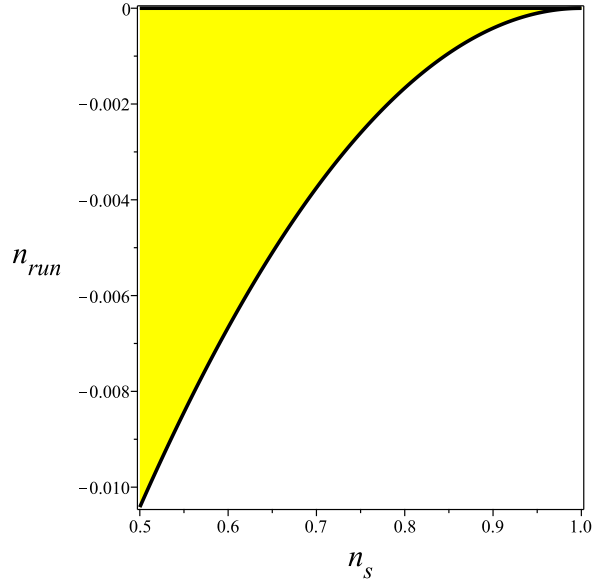}
\caption{The allowed yellow region in $(n_{run}-n_s)$ plane for our model lies between the curves $n_{run}=0$ and $n_{run}=-\frac{1}{24}(1-n_s)^2$.}\label{fig2}
\end{figure}

In a similar manner we can study the trajectories in $(n_{run}-n_s)$ plane. Using Eqs.(\ref{ns}) and (\ref{nrun}) we come to the following relation
\begin{equation}\label{nrun-ns}
n_{run}=\frac{-\frac{\alpha}{H}(1-n_s)^2}{\left(2+3\frac{\alpha}{H}\right)^2}
\end{equation}
Clearly, for the case $\alpha\ll H$ we come to $n_{run}\approx0$. This can be considered as one boundary trajectory in $(n_{run}-n_s)$ plane. Also, for $\alpha\gg H$, we find $n_s\approx1$, again. To find another boundary trajectory to exert in $(n_{run}-n_s)$ plane as a more powerful constraint we return to Eq.(\ref{nrun}). We find that $n_{run}\approx0$, for the case $\alpha\gg H$, as well. But during inflation and along with changing $H$, the parameter $n_{run}$ varies, too. To calculate the maximum possible variation of $n_{run}$ versus $n_s$, we differentiate Eq.(\ref{nrun-ns}) with respect to $\frac{\alpha}{H}$ and set it to zero. The result is $\frac{\alpha}{H}=\frac{2}{3}$ which yields $n_{run}=-\frac{1}{24}(1-n_s)^2$ as the second boundary trajectory. The allowed region in $(n_{run}-n_s)$ plane has been shown in FIG.\ref{fig2}. It means that with attention to Eqs.(\ref{nsN}) and (\ref{nrunN}), for all set of parameters $(A,\lambda,\alpha)$ the trajectories fall within this region.

Now, we compare our results with observations. To this aim we refer to Planck 2018 outcomes \cite{Akrami},\cite{Aghanim}. One of the most significant results from Planck 2018 is that the spectrum of scalar fluctuations is red-tilted. By red tilt we mean $n_s<1$. FIG.\ref{fig1} clearly illustrates that in our model, $n_s$ never gain the values bigger than one. In this sense the model under consideration is consistent with the latest observations. In comparison with the results in \cite{Ravanpak} for a tachyon logamediate inflation and also in \cite{Ravanpak2} for a tachyon DGP intermediate inflation, that demonstrated blue tilt for some specific choices of model parameters, this is a fascinating consequence. We would like to note that the Planck collaboration have investigated several one- and two-parameter extensions of the base-$\Lambda$CDM model, with varying $r$, $n_{run}$ and some other parameters such as the sum of neutrino masses, spatial curvature and so on. Surprisingly, all of them support the red-tilted scalar spectrum though their confidence regions differ slightly. We have to note that though our model is compatible with a red-tilted scalar spectrum, but it shows higher values for the scalar spectral index, $n_s$, than the observational results. For example, it has indicated in \cite{Aghanim} that in a $\Lambda$CDM+$r$+$n_{run}$ model, $n_s = 0.9647\pm0.0044$ at 68\% confidence limit.

Meanwhile, by combining Planck temperature, low-$\ell$ polarization and lensing, the authors in \cite{Akrami}, have obtained a new constraint on the tensor-to-scalar ratio as $r_{0.002}<0.10$ which has been demonstrated as a horizontal black dashed line in FIG.\ref{fig1} which crosses the allowed region in $(r-n_s)$ plane of our model. So, as we mentioned earlier, all the possible trajectories in $(r-n_s)$ plane related to various combination of model parameters, will most likely go through this line and enter the constrained region over time. We have shown this behavior for two different choices in FIG.\ref{fig1}. The red (left) curve is related to ($\lambda$=10, $A=10^{-11}$) and the blue (right) curve is related to ($\lambda$=20, $A=4.7\times10^{-28}$). For both of them we have considered $\alpha=2\times10^{-12}$. Also, the red diamond and the red circle relates to the number of $e$-fold parameter $N=100$ and $N=300$, respectively while the blue diamond and the blue circle indicates $N=60$ and $N=150$, respectively. So, it is clear that they both go across the constraint line at a given $N$ and come into the region $r<0.1$. Although both these curves enter into this constrained region for nearly large values of $N$ which is not admissible, but one might find some other pairs of ($\lambda$, $A$) that satisfy the constraints on $r$ for appropriate values of $N$ and show more consistency of the model.

Another important result of Planck 2018 is that the value of running in the scalar spectral index is close to zero. As we mentioned earlier, a zero running implies that $n_s$, is independent of scale. In both \cite{Akrami} and \cite{Aghanim}, the authors report $n_{run}=-0.0045\pm0.0067$ at 68$\%$ confidence limit for the $\Lambda$CDM+$n_{run}$ model, whereas in the presence of tensor modes this value does not change more than $n_{run}=-0.0085\pm0.0073$ \cite{Aghanim}. Altogether, $n_{run}$ has a very small negative value, at the most. This is consistent with our results in FIG.{\ref{fig2}}.

\begin{figure}[h]
\centering
\includegraphics[width=8cm]{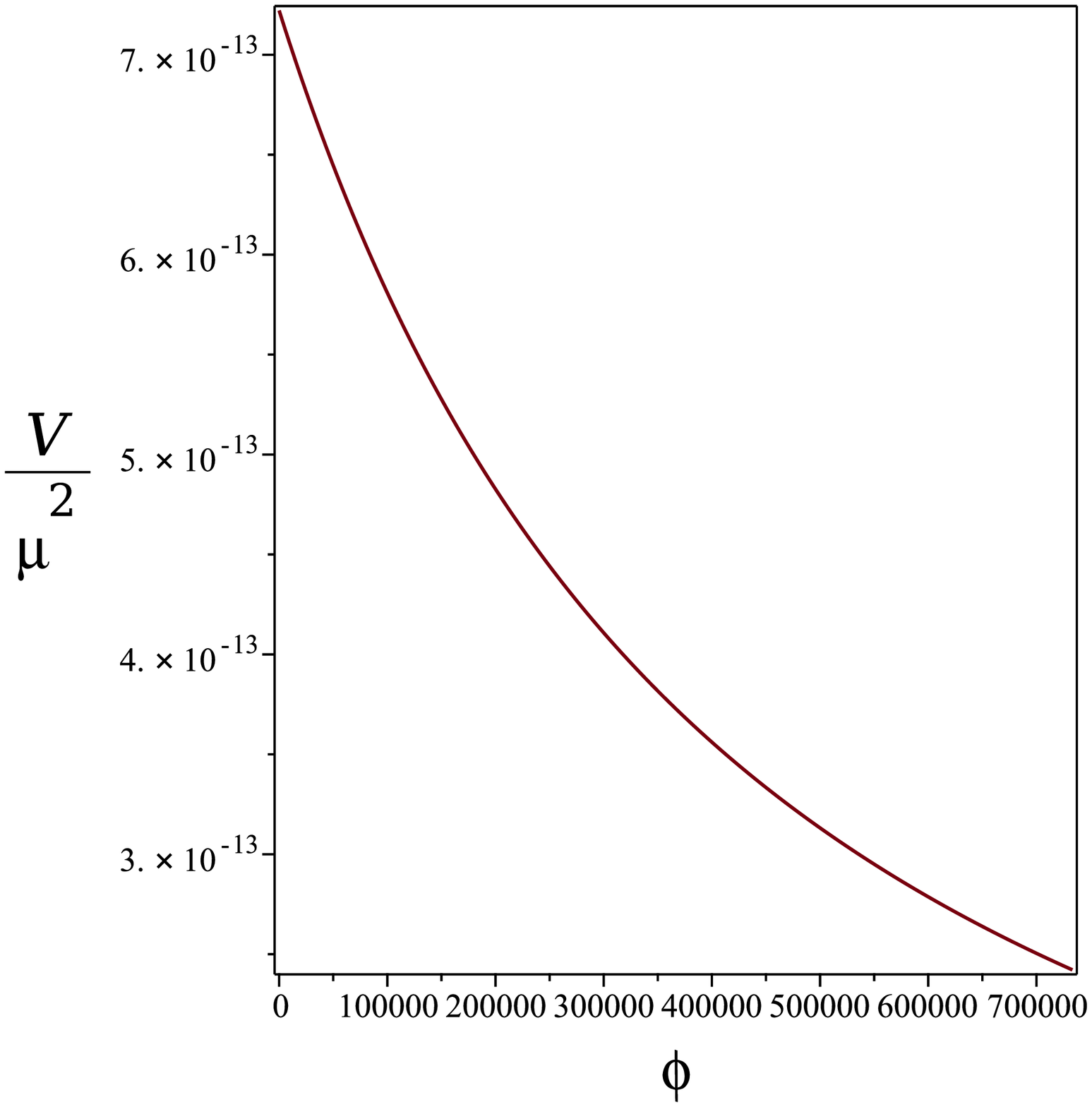}
\includegraphics[width=8cm]{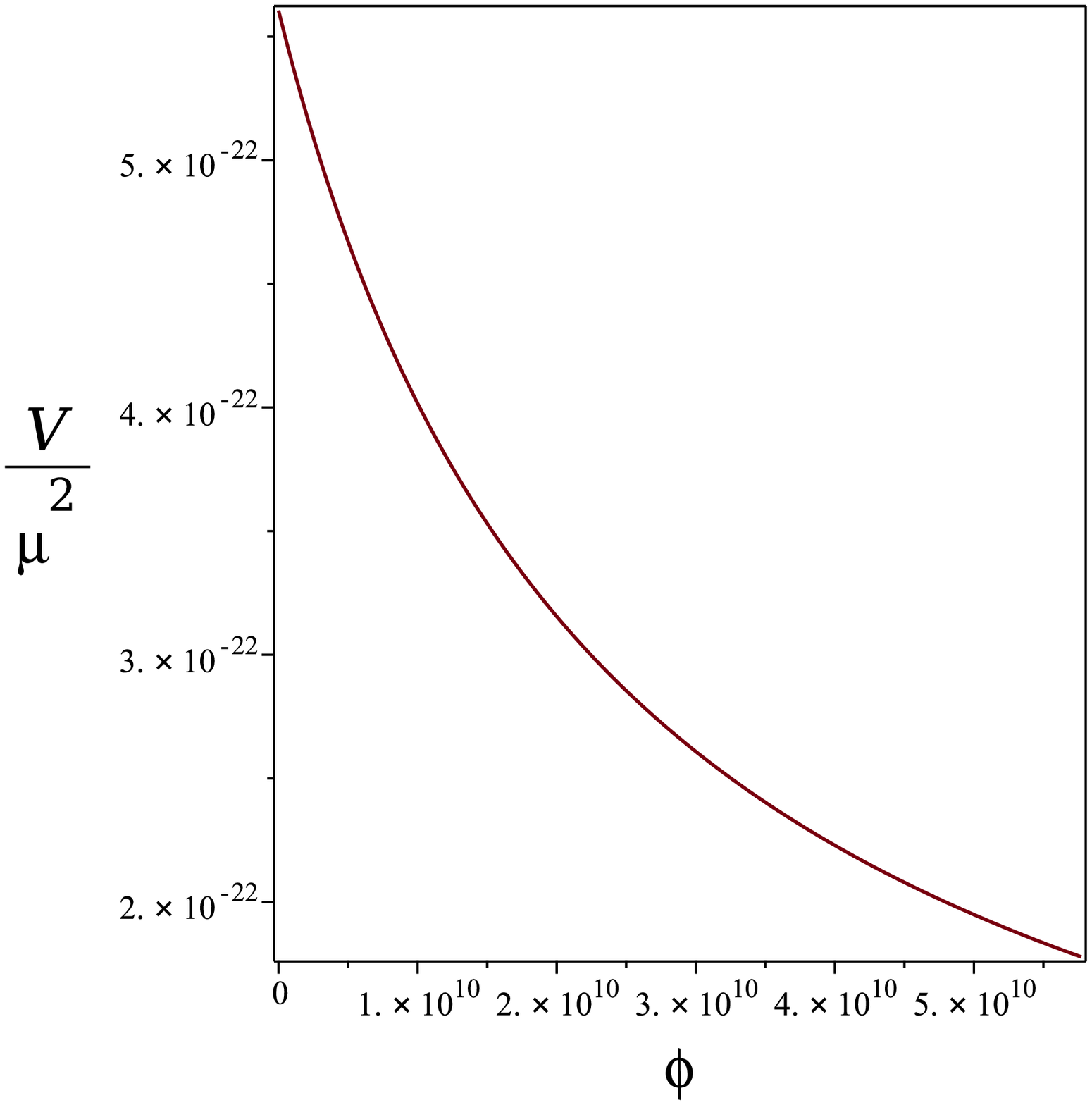}
\caption{The behavior of $V(\phi)/\mu^2$ for ($\lambda$=10, $A=10^{-11}$) (the left) and ($\lambda$=20, $A=4.7\times10^{-28}$) (the right) with $\alpha=2\times10^{-12}$.}\label{fig3}
\end{figure}

To check how realistic is the model under consideration we can plot $V(\phi)$ for different values of model parameters using Eqs.(\ref{phi}) and (\ref{SReffectivepotential}). FIG.{\ref{fig3}} left, is related to ($\lambda$=10, $A=10^{-11}$) and FIG.{\ref{fig3}} right, to ($\lambda$=20, $A=4.7\times10^{-28}$), while in both of them we have assumed $\alpha=2\times10^{-12}$. As we mentioned in Introduction, the tachyonic potential is an ever-decreasing function. We see that in our model $V(\phi)/\mu^2$ obeys such a behavior. In plotting both the graphs we have considered $t_b$ as the required initial time.

\section{conclusion}\label{sec5}

We have analyzed the logamediate inflationary scenario in the DGP braneworld cosmology in the presence of a tachyon scalar field. Applying slow-roll conditions and in high energy
regime we found effective Friedmann equation and the slow-roll quantities in terms of our model parameters. Furthermore, explicit expressions for the scalar spectral index $n_s$, its running $n_{run}$ and also tensor-to-scalar ratio $r$ are derived.

To compare our findings with observations we utilized Planck 2018 results in \cite{Akrami} and \cite{Aghanim}, in which different extensions of the basic $\Lambda$CDM cosmological model have been investigated. We understood that in our model the spectrum of scalar fluctuations is always red-tilted that means $n_s<1$. Also, we found that $n_{run}\simeq0$ so that in the case of $n_{run}\neq0$, it has a very small negative value. We realized that in all of the situations there are good agreements between our results and those of Planck 2018 except for the value of $n_s$ and also for the value of the number of $e$-folds for which the trajectories in $(r-n_s)$ plane cross the latest constraint on tensor-to-scalar ratio parameter, $r=0.1$.

The work in this article is just a part of the whole framework of the inflationary paradigm. The inclusion of the reheating epoch and transition to the standard cosmology may lead to new constraints on our model parameters.


\begin{thebibliography}{}

\bibitem{Guth} A. Guth, Phys. Rev. D 23, 347 (1981).
\bibitem{Albrecht} A. Albrecht and P. J. Steinhardt, Phys. Rev. Lett. 48, 1220 (1982).
\bibitem{Liddle} A. R. Liddle and D. H. Lyth, \textit{Cosmological Inflation and Large-Scale Structure}, Cambridge University Press, Cambridge (2000).
\bibitem{Dodelson} S. Dodelson, \textit{Modern Cosmology}, Academic Press, San Diego (2003).
\bibitem{Dunkley} J. Dunkley et al., Astrophys. J. Suppl. Ser. 180, 306 (2009).
\bibitem{Hinshaw} G. Hinshaw et al., Astrophys. J. Suppl. Ser. 180, 225 (2009).
\bibitem{Barrow} J. D. Barrow, Phys. Rev. D 51, 2729 (1995).
\bibitem{Barrow2} J. D. Barrow, Classical Quantum Gravity 13, 2965 (1996).
\bibitem{Farajollahi} H. Farajollahi and A. Ravanpak, Can. J. Phys. 89, 1015 (2011).
\bibitem{Farajollahi2} H. Farajollahi and A. Ravanpak, Can. J. Phys. 88, 939 (2010).
\bibitem{Sen} A. Sen, J. High Energy Phys. 04, 048 (2002).
\bibitem{Sen2} A. Sen, J. High Energy Phys. 07, 065 (2002).
\bibitem{Sami} M. Sami, P. Chingangbam and T. Qureshi, Phys. Rev. D 66, 043530 (2002).
\bibitem{Picon} C. Armendariz-Picon, T. Damour and V. Mukhanov, Phys. Lett. B 458, 209 (1999).
\bibitem{Ravanpak} A. Ravanpak and F. Salmeh, Phys. Rev. D 89, 063504 (2014).
\bibitem{Setare} M. Setare, A. Ravanpak and H. Farajollahi, Gravitat. Cosmol. 24, 52 (2018).
\bibitem{Farajollahi3} H. Farajollahi and A. Ravanpak, Phys. Rev. D 84, 084017 (2011).
\bibitem{Campo} S. del Campo and R. Herrera, Phys. Lett. B 670, 266 (2009).
\bibitem{Sami2} M. Sami, Mod. Phys. Lett. A 18, 691 (2003).
\bibitem{Maartens} R. Maartens, D. Wands, B. A. Bassett, and I. P. C. Heard, Phys. Rev. D 62, 041301 (2000).
\bibitem{Kamali} V. Kamali and E. Navaee Nik, Eur. Phys. J. C 77, 449 (2017).
\bibitem{Dvali} D. Dvali, G. Gabadadze and M. Porrati, Phys. Lett. B 485, 208 (2000).
\bibitem{Herrera} R. Herrera, N. Videla and M. Olivares, Phys. Rev. D 90, 103502 (2014).
\bibitem{Herrera2} R. Herrera, M. Olivares and N. Videla, Eur. Phys. J. C 73, 2475 (2013).
\bibitem{Nozari} K. Nozari and B. Fazlpour, J. Cosmol. Astropart. Phys. 0711, 006 (2007).
\bibitem{Campo2} S. del Campo and R. Herrera, Phys. Lett. B 653, 122 (2007).
\bibitem{Cai} R. G. Cai and H. Zhang, J. Cosmol. Astropart. Phys. 0408, 017 (2004).
\bibitem{Ravanpak2} A. Ravanpak, H. Farajollahi and G. F. Fadakar, Astrophys. Space Sci. 361, 43 (2016).
\bibitem{Lopez} M. Bouhmadi-Lopez and L. Chimento, Phys. Rev. D 82, 103506 (2010).
\bibitem{Maeda} K. Maeda, S. Mizuno and T. Torii, Phys. Rev. D 68, 024033 (2003).
\bibitem{Barrow3} J. D. Barrow and N. J. Nunes, Phys. Rev. D 76, 043501 (2007).
\bibitem{Hwang} J. C. Hwang and H. Noh, Phys. Rev. D 66, 084009 (2002).
\bibitem{Wands} D. Wands, K. A. Malik, D. H. Lyth and A. R. Liddle, Phys. Rev. D 62, 043527 (2000).
\bibitem{Campo3} S. del Campo, R. Herrera and A. Toloza, Phys. Rev. D 79, 083507 (2009).
\bibitem{Lopez2} M. Bouhmadi-Lopez, R. Maartens and D. Wands, Phys. Rev. D 70, 123519 (2004).
\bibitem{Lidsey} J. E. Lidsey, A. R. Liddle, E. W. Kolb, E. J. Copeland, T. Barreiro and M. Abney, Rev. Mod. Phys. 69, 373 (1997)
\bibitem{Akrami} Y. Akrami, et al. (Planck Collaboration), Astron. Astrophys. 641, A10 (2020).
\bibitem{Aghanim} N. Aghanim, et al. (Planck Collaboration), Astron. Astrophys. 641, A6 (2020).

\end{thebibliography}
\end{document}